\documentclass[reprint,
superscriptaddress,
amsmath,amssymb,aps,showkeys,showpacs,
twoside,final,secnumarabic,
nofootinbib]{revtex4-2}

\usepackage[paperwidth=205mm,paperheight=290mm,top=17mm,bottom=25mm,
inner=17mm,outer=17mm,
twoside]{geometry}

\usepackage{cmap} 
\usepackage[T1,T2A]{fontenc}
\usepackage[utf8]{inputenc}
\usepackage[russian,english]{babel}
\usepackage{color}
\usepackage{graphicx}
\usepackage{dcolumn}
\usepackage{bm} 
\usepackage[unicode=true,colorlinks=true,linkcolor=magenta, urlcolor=blue, citecolor = blue,breaklinks]{hyperref}
\usepackage{multirow}
\usepackage{url}
\usepackage{breakurl}
\usepackage{graphicx}
\usepackage{placeins}
\usepackage{relsize}

\DeclareGraphicsExtensions{.eps}

\newcount\issue
\newcount\Vol
\newcount\numb
\usepackage{fancyhdr} 
\def\Vol{\textbf{78}}
\def\numb{x}
\def\babar{\mbox{\slshape B\kern-0.1em{\smaller A}\kern-0.1em B\kern-0.1em{\smaller A\kern-0.2em R}}}

\begin{document}

\title{JOURNAL SECTION OR CONFERENCE SECTION\\[20pt]
Searches for invisible new particles
at Belle~II} 

\def\addressa{INFN Sezione di Pisa, I-56127 Pisa, Italy}

\author{\firstname{L.}~\surname{Corona}~\surname{on behalf of the Belle~II Collaboration}}
\email[E-mail: ]{luigi.corona@pi.infn.it }
\affiliation{\addressa}

\received{xx.xx.2023}
\revised{xx.xx.2023}
\accepted{xx.xx.2023}

\begin{abstract}
Belle~II has unique sensitivity for a broad class of models postulating the existence of dark matter particles with masses in the MeV--GeV range. We present recent world-leading results from Belle~II searches for several non-SM particles.
These include production of $Z'$ bosons, axion-like particles, and dark scalars in association with two muons in $e^+e^-$ collisions; long-lived (pseudo)scalars produced in decays of \textit{B}-mesons; and invisible particles produced in decays of $\tau$ leptons. 
\end{abstract}

\par
\keywords{Dark Matter, Dark Sector, Belle~II, $Z'$, axion-like particle, dark scalar, leptophilic, muonphilic  \\[5pt]}

\maketitle
\thispagestyle{fancy}


\section{Introduction}\label{intro}

Several astrophysical observations suggest the existence of dark matter (DM), a component of matter that does not interact through strong or electromagnetic forces. Although DM constitutes approximately 85\% of the total matter in our Universe, its nature remains unknown. 
Dark Matter is one of the most compelling phenomena in support for physics beyond the Standard Model (SM). 

The lack of evidence of non-SM physics at the electroweak scale leads to hypothesize 
sub-GeV DM particles feebly interacting with SM particles through non-SM mediators. 
Sub-GeV DM and the non-SM mediators belong to the dark sector, and efforts to detect them 
have been actively pursued at beam dump and high-intensity frontier experiments. 

Belle~II~\cite{B2TIP, B2TDR} is a high-intensity frontier experiment that operates at the SuperKEKB $e^+e^-$ asymmetric-energy collider~\cite{SKEKB}.  
During the first data taking run (2019--2022), Belle~II collected a sample of $e^+e^-$ collision data corresponding to 424~fb$^{-1}$ of integrated luminosity. 
Thanks to the excellent reconstruction capabilities for low multiplicity and missing energy signatures, and dedicated triggers, Belle~II has a unique or world-leading sensitivity to dark sector~\cite{snow}. 

\section{\label{sec:results} Recent Dark Sector results at Belle~II}

\subsection{\label{sec:zpinvisible}Search for an invisible \texorpdfstring{$Z'$}{}}

\begin{figure}[!ht]
\includegraphics[width=0.98\linewidth]{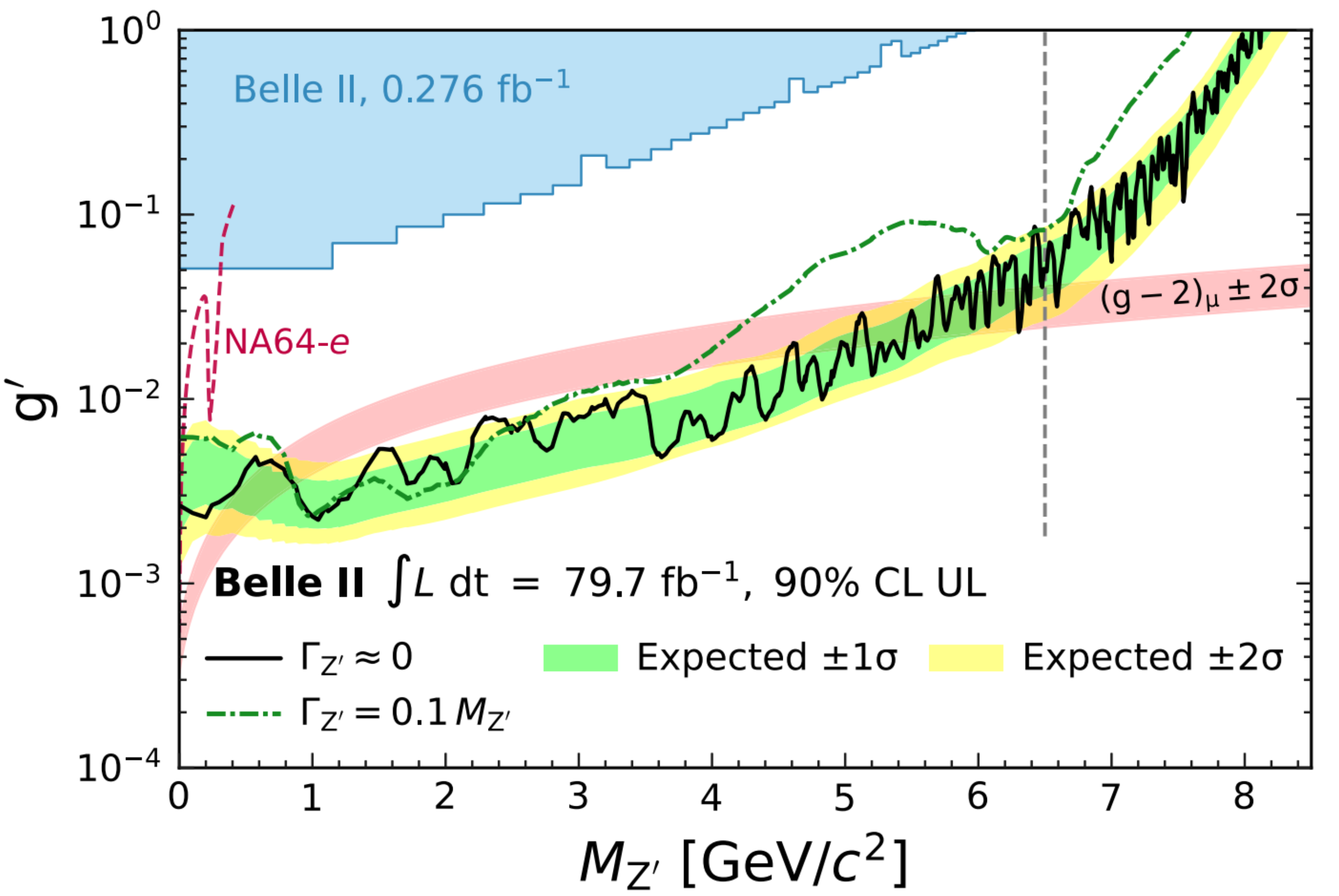}
\caption{\label{fig:zpinv} Observed 90\%~C.L. upper limits and corresponding expected limits on the $g'$ coupling as a function of the $Z'$ mass, assuming $\mathcal{B}(Z' \to \mathrm{inv.}) = 100\%$.}
\end{figure}

The $L_{\mu} - L_{\tau}$ model~\cite{Zp1, Zp2, Zp3} introduces a light gauge boson, $Z’$, that violates lepton-flavor universality while conserving the difference between $\mu$ and $\tau$ lepton numbers.  
We search for the invisible decay of the $Z'$ 
through the process $e^+e^- \to \mu^+\mu^- Z' (\to \mathrm{inv.})$, where the $Z'$ is radiated off one of the muons.  
The $Z'$ could decay invisibly to SM neutrinos, with a branching fraction of $\mathcal{B}(Z' \to \mathrm{inv.}) \sim 33\%$, or to kinematically accessible DM candidates 
with $\mathcal{B}(Z' \to \mathrm{inv.}) = 100\%$.  
A signal would appear as a narrow enhancement in the recoil mass against the two final-state muons, in events where nothing else is detected. The main backgrounds are QED radiative di-lepton and four-lepton final states. 
The backgrounds are suppressed using a neural-network trained simultaneously for all $Z'$ masses~\cite{PunziNET}, and fed with kinematic variables sensitive to the origin of the missing energy: in the signal, the $Z'$ is produced as final-state radiation (FSR); in the background, the missing energy is due to neutrinos or undetected particles. 
From 2D template fits to the recoil mass squared, in bins of recoil polar angle, we do not observe any significant excess in 79.7~fb$^{-1}$ of data, and we set 90\%~C.L. upper limits on the coupling of the $L_\mu - L_\tau$ model, $g'$, as a function of the $Z'$ mass, $M_{Z'}$. We exclude the region favored by the $(g-2)_{\mu}$ anomaly~\cite{g2}, which could be explained by the $L_\mu - L_\tau$ model, in the mass range $0.8<M_{Z'}<5$~GeV/$c^2$ for the fully invisible $L_\mu - L_\tau$ model (Fig. \ref{fig:zpinv})~\cite{Zpinv2020, Zpinv2023}.

\subsection{\label{sec:xtautau}Search for \texorpdfstring{$e^+e^- \to \mu^+\mu^-X(\to \tau^+\tau^-)$}{}}

We search for a $X \to \tau^+\tau^-$ resonance, where $X$ could be a $Z'$, a leptophilic dark scalar $S$, or an axion-like particle (ALP), in $e^+e^- \to \mu^+\mu^-\tau^+\tau^-$ events,  with $\tau$ decaying to one charged particle. The $S$ is an hypothetical particle that couples preferentially to charged leptons through Yukawa-like couplings~\cite{lds}. Axion-like particles are pseudo-scalars that appear in many SM extensions~\cite{alp2017, alp2022}. 

\begin{figure}[b] 
  \centering
    \includegraphics[width=0.98\linewidth]{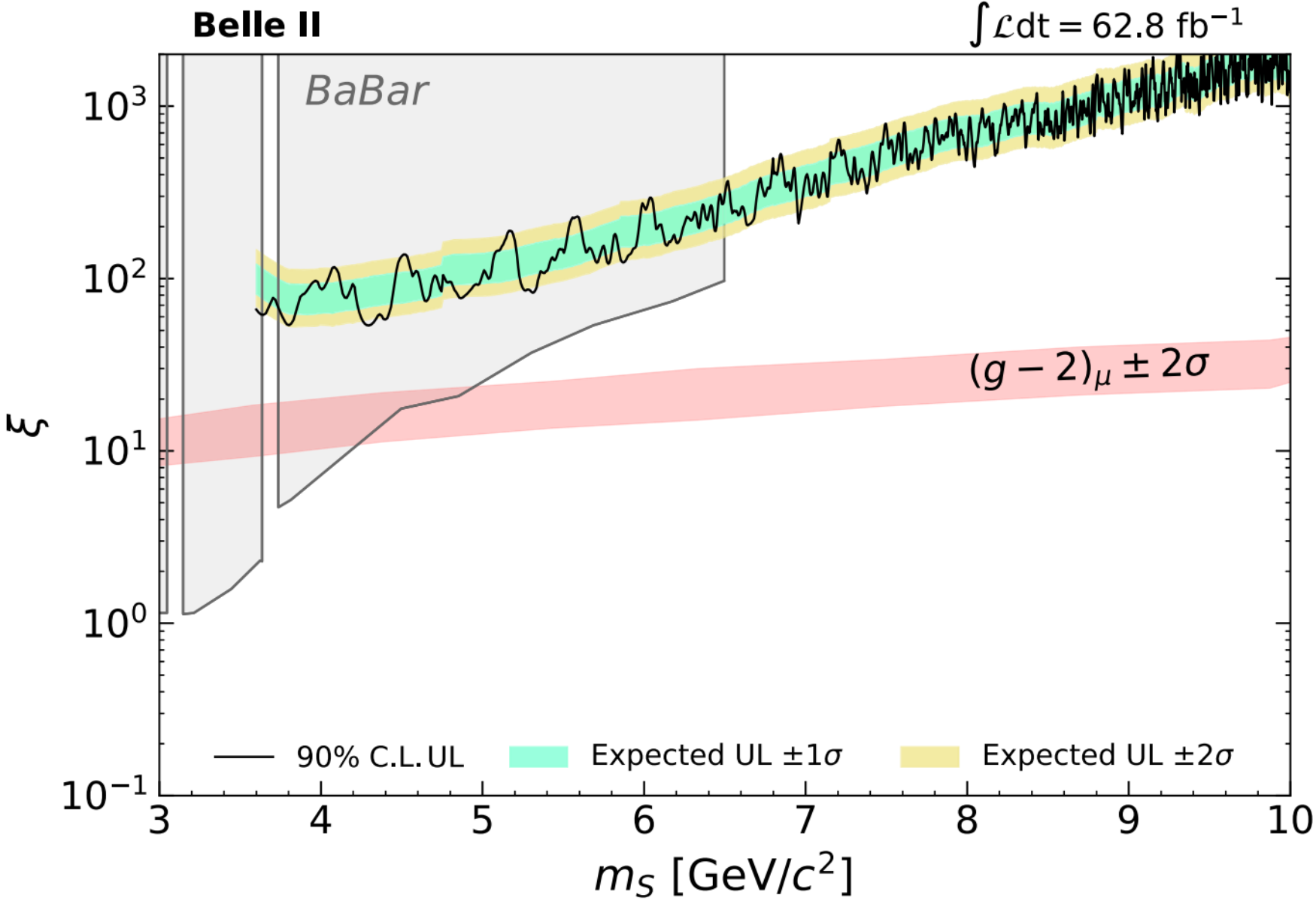}
    \includegraphics[width=0.98\linewidth]{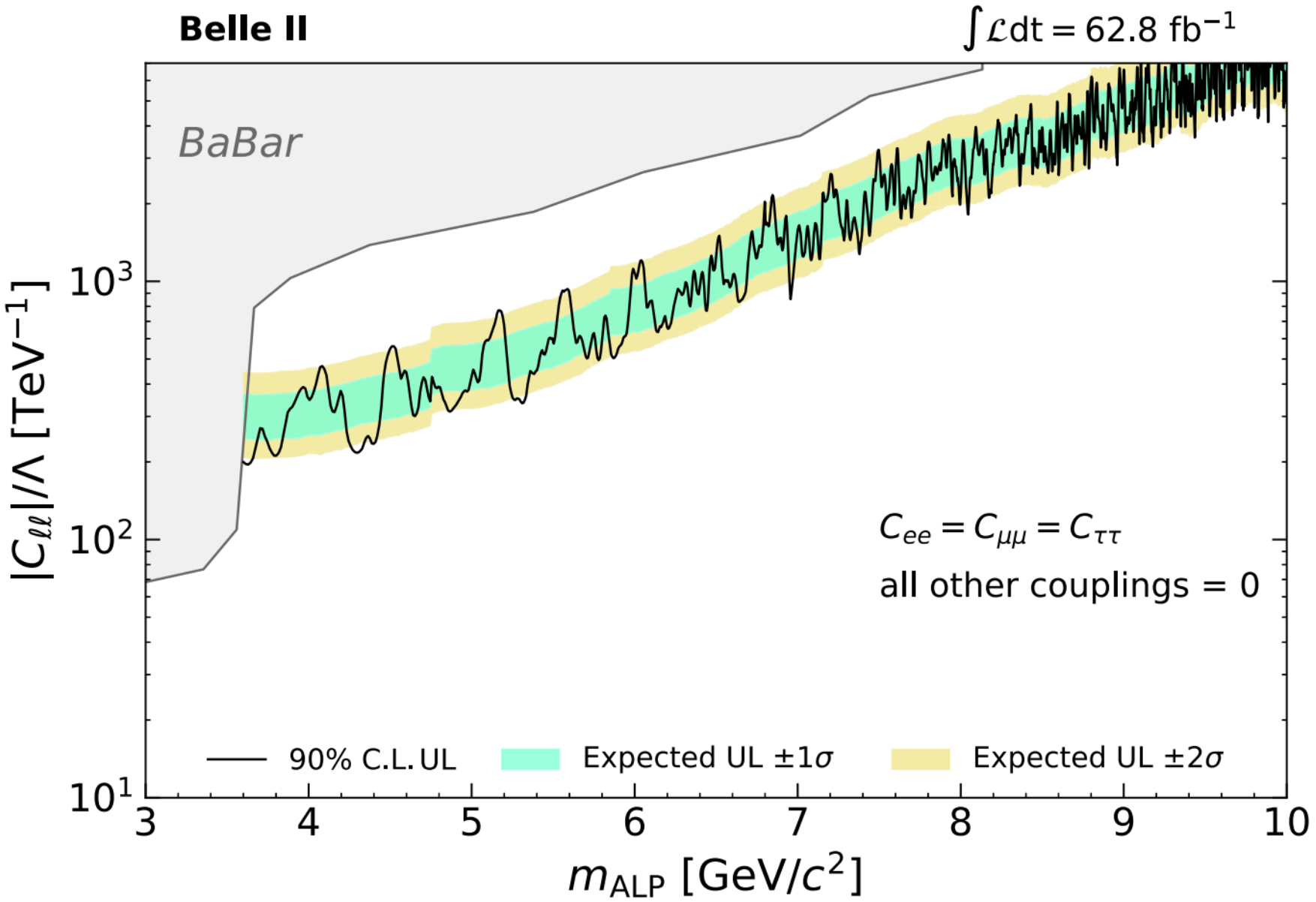}
    \caption{Observed 90\%~C.L. upper limits (UL) and corresponding expected limits as a function of the mass on ({\it top}) the leptophilic scalar coupling $\xi$, and on ({\it bottom}) 
    the ALP-lepton coupling 
    $|C_{\ell\ell}|/\Lambda$.} 
    \label{fig:xtautau}
\end{figure}

Similarly to the $Z' \to \rm{inv.}$ analysis, we search for a narrow enhancement in the recoil mass against two oppositely charged muons, in four-track events with zero net charge.  
Standard Model backgrounds are suppressed with eight neural-networks fed with kinematic variables sensitive to the $X$-production mechanism as FSR off one of the two muons, and trained in different $X$-mass regions. From extended maximum likelihood fits to the recoil mass distribution, we do not observe any significant excess in 62.8~fb$^{-1}$ of data. We derive world-leading 90\%~C.L. upper limits on the $S$-coupling $\xi$ for $m_S > 6.5$ GeV/$c^2$, and on the ALP-lepton coupling $|C_{\ell\ell}|/\Lambda$, assuming equal ALP-couplings to the three lepton families and zero couplings to all other particles (Fig. \ref{fig:xtautau})~\cite{xtautau}.

\subsection{\label{sec:xmumu}Search for \texorpdfstring{$e^+e^- \to \mu^+\mu^-X(\to \mu^+\mu^-)$}{}}

\begin{figure}[!hb] 
  \centering
    \includegraphics[width=0.98\linewidth]{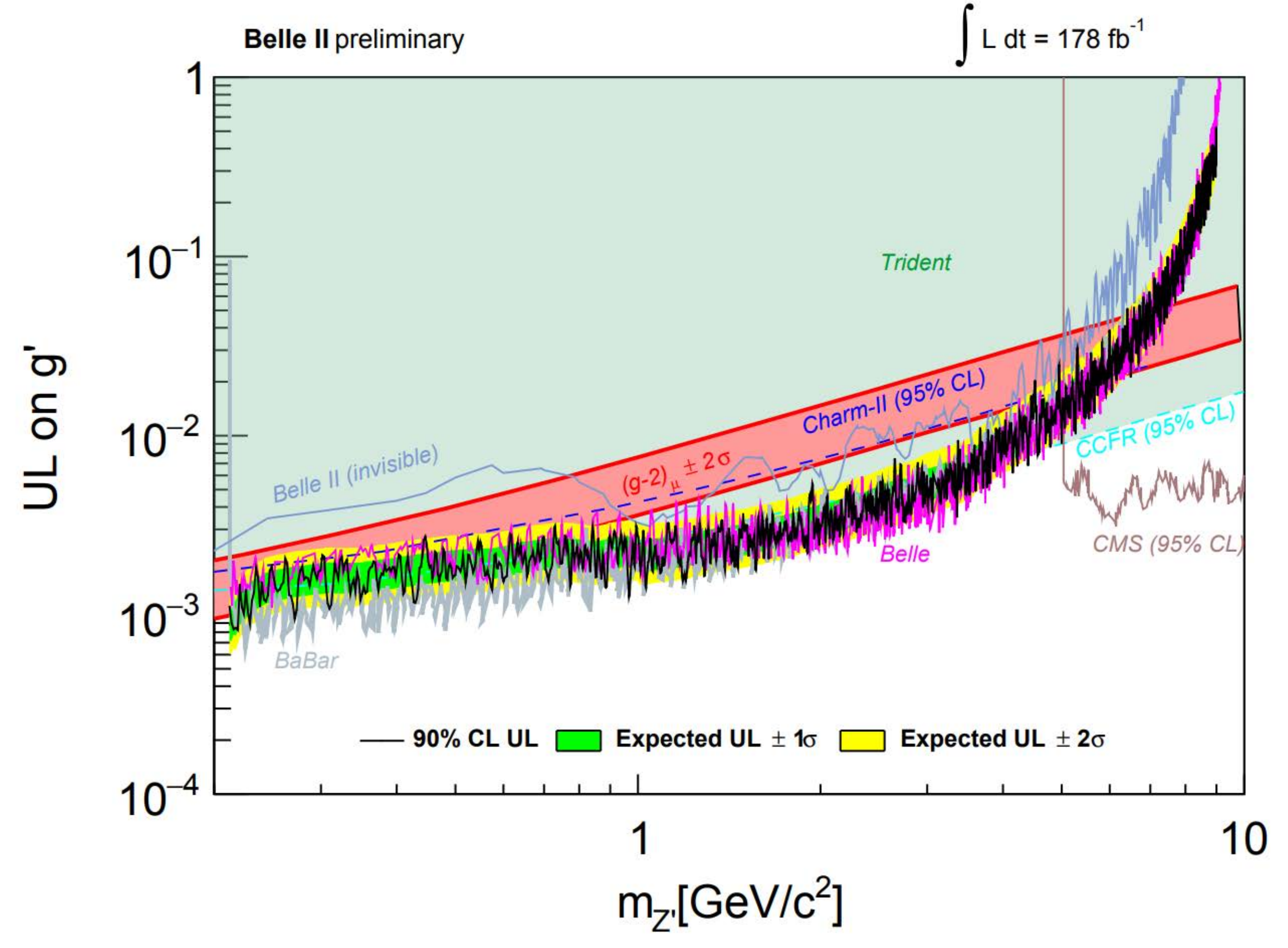}
    \includegraphics[width=0.98\linewidth]{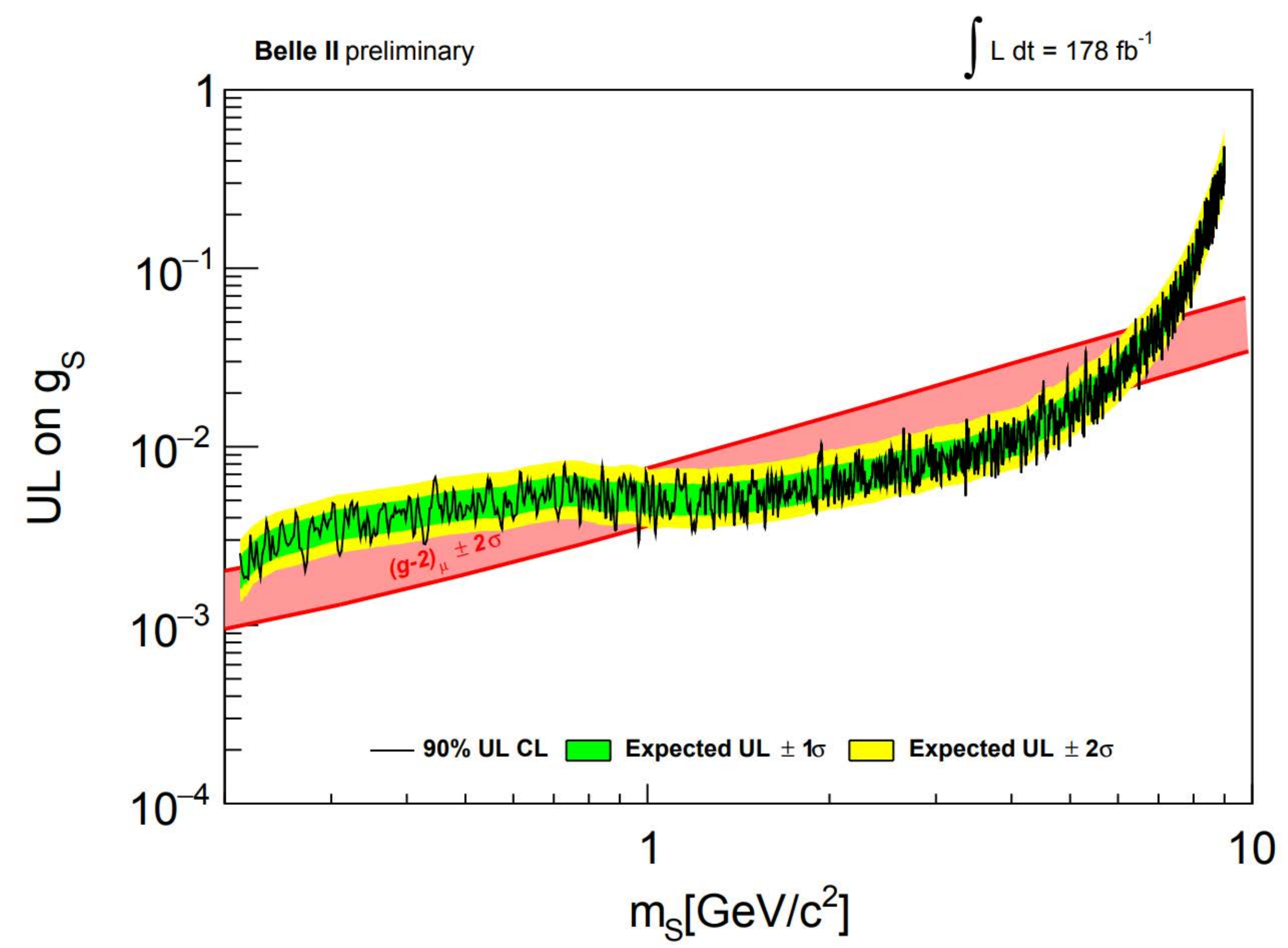}
    \caption{Observed 90\%~C.L. upper limits and corresponding expected limits as a function of the mass on ({\it top}) the $g'$ coupling of the $L_{\mu} - L_{\tau}$ model, and on ({\it bottom})
    the muonphilic dark scalar model.} 
    \label{fig:xmumu}
\end{figure}

We search for a $X \to \mu^+\mu^-$ resonance in $e^+e^- \to \mu^+\mu^-\mu^+\mu^-$ events as a narrow enhancement in the dimuon mass distribution in four-track events with zero net charge and no extra-energy. The dominant background is the 
SM four-muon final-state process. 
Background is suppressed applying five neural-networks fed with kinematic variables sensitive to the $X$-production mechanism as FSR off one of the two muons, and to the presence of a resonance in both the candidate and the recoil muon pairs, and trained in different $X$-mass ranges.  
From extended maximum likelihood fits to the dimuon mass distribution, we do not observe any significant excess in 178~fb$^{-1}$ of data, and we set 90\%~C.L. upper limits on the cross section of the process. 
We interpret the results obtained on the cross section as 90\%~C.L. limits on the $g'$ coupling of the $L_{\mu} - L_{\tau}$ model, and on the coupling of a muonphilic  
scalar $S$ with muons~\cite{mds}. Despite the small data-set used, we obtain similar results with the existing limits on $g'$ from \babar~\cite{babar} and Belle~\cite{belle}, which performed the analysis with 514~fb$^{-1}$ and 643~fb$^{-1}$ respectively. 
We set the first limits for the muonphilic scalar model from a dedicated search (Fig. \ref{fig:xmumu}).

\subsection{\label{sec:llp}Search for a long-lived (pseudo)scalar in \texorpdfstring{$b \to s$}{} transitions}

Some extensions of the SM introduce a new light scalar $S$ that may give mass to DM particles. The scalar $S$ would mix with the SM Higgs boson through a mixing angle $\theta_S$, and would be naturally long-lived for small values of $\theta_S$. 

\begin{figure}[!ht]
\includegraphics[width=0.98\linewidth]{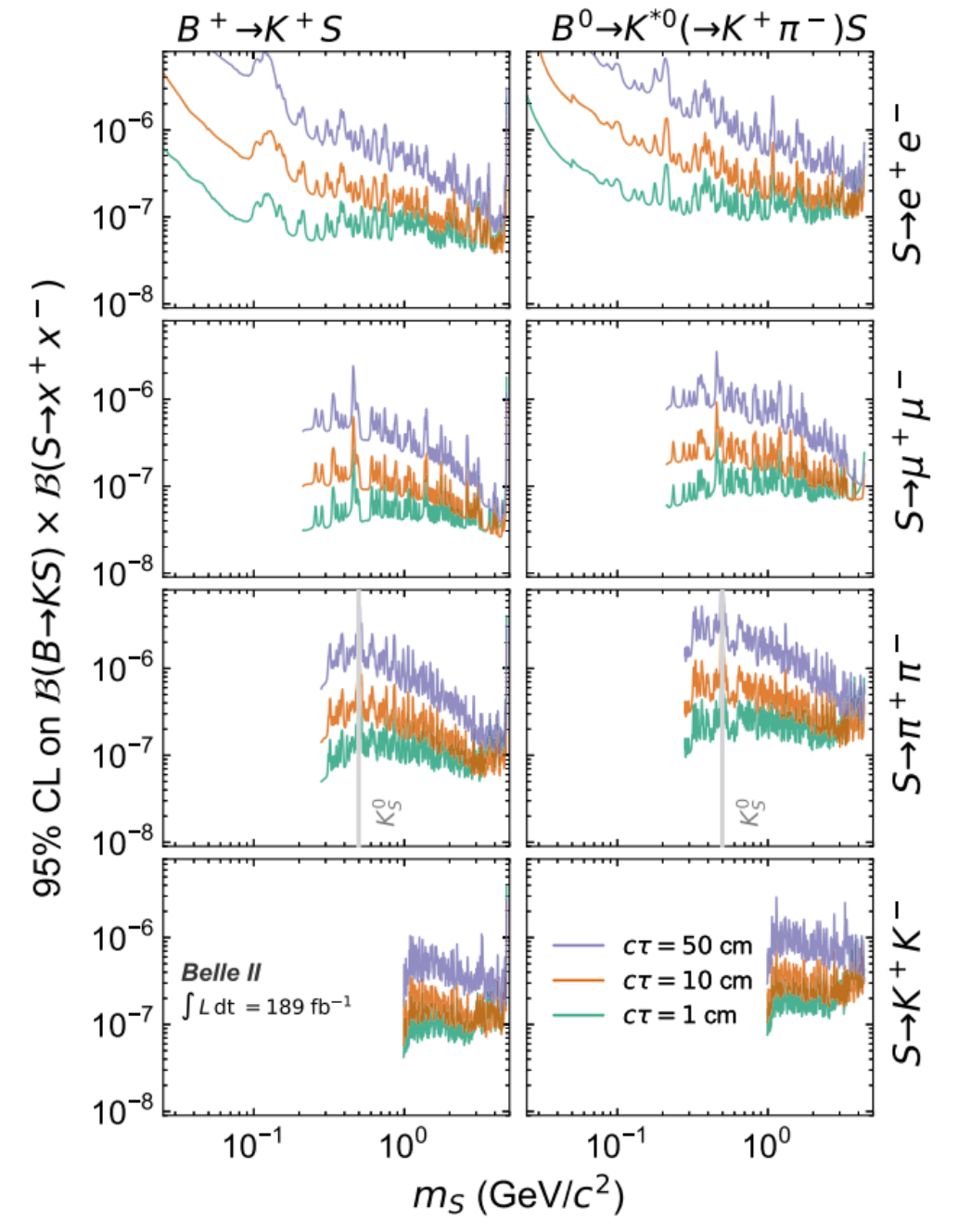}
\caption{\label{fig:llp} Observed 95\% C.L. on $\mathcal{B}(B \to KS)\times \mathcal{B}(S\to x^+x^-)$ as a function of the scalar mass $m_S$ for different lifetimes $c\tau$.}
\end{figure}

We search for $B^0 \to K^{*0}(\to K^+\pi^-)S$ and $B^+ \to K^{+}S$ events, with $S \to x^+x^-$ ($x = e,\mu,\pi,K$) forming a decay vertex displaced from the \textit{B} decay vertex.  
The signal yield is extracted through extended maximum likelihood fits to the reduced invariant mass of $S$, $m^{r}_{S\to xx} = \sqrt{M^2_{S \to xx} - 4m^2_{x}}$, in order to improve the modeling of the signal width close to the kinematic thresholds. Main background components are the combinatorial
$e^+e^- \to q\bar{q}$, suppressed by requiring a kinematics similar to \textit{B}-meson expectations; 
$B \to K K_S(\to \pi^+\pi^-)$, vetoed; and $B\to K x^+x^-$ decays without intermediate long-lived particles decaying to $x^+x^-$, suppressed by tightening the displacement selections. we do not observe any significant excess in 189~fb$^{-1}$ of data, and we set the first model-independent limits at 95\% C.L. on $\mathcal{B}(B \to KS)\times \mathcal{B}(S\to x^+x^-)$ as a function of the scalar mass $m_S$ for different $S$-lifetimes (Fig. \ref{fig:llp})~\cite{btos}. 

\newpage
\subsection{\label{sec:alpha}Search for the  \texorpdfstring{$\tau \to \ell\alpha$}{} decay}

Charged-lepton flavour violation (LFV) is allowed in various extensions of the SM, however it has never been observed. In these extensions, the LFV processes could be mediated by a new hypothetical $\alpha$ boson. We search for an invisible $\alpha$ produced in the $\tau \to \ell\alpha$ decay, with $\ell = e,\mu$, in $e^+e^- \to \tau^+\tau^-$ events. In the center-of-mass frame, $\tau$ pairs are produced back-to-back so that the decay products of each $\tau$ lepton are contained in two separate hemispheres. The \textit{tag} hemisphere contains three charged hadrons from $\tau_{\rm{tag}}^- \to h^-h^+h^-\nu_{\tau}$, with $h=\pi$, $K$, while the \textit{signal} hemisphere contains only one charged lepton from the $\tau^-_{\rm{sig}}\to\ell^-\alpha$ decay.

\begin{figure}[!ht] 
  \centering
    \includegraphics[width=0.98\linewidth]{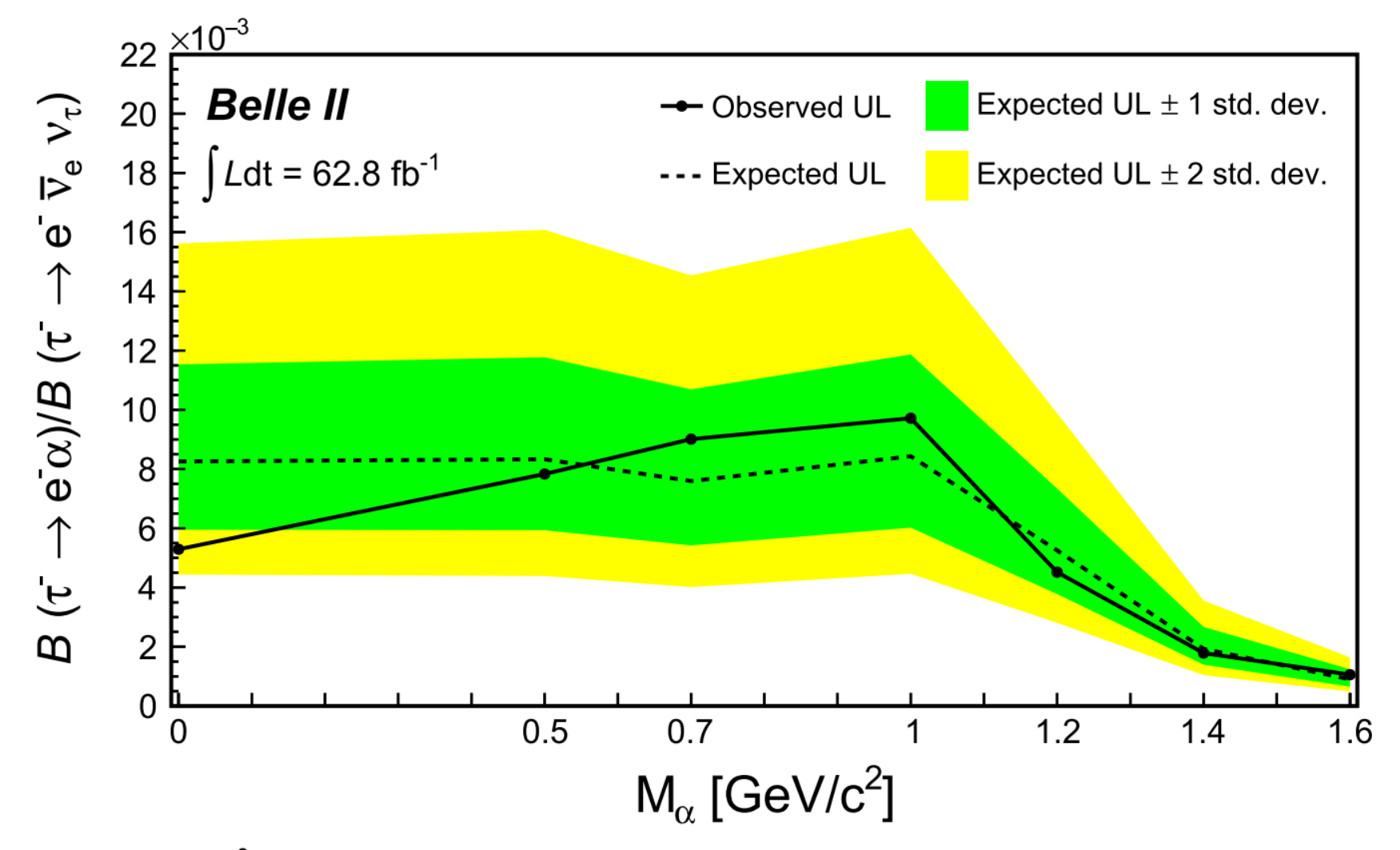}
    \caption{Upper limits at 95\% C.L. 
    on the ratio $\mathcal{B}(\tau \to e\alpha)/\mathcal{B}(\tau \to e\bar{\nu}_{e}\nu_{\tau})$.} 
    \label{fig:tau}
\end{figure}

For this analysis, $\tau \to \ell\nu_{\tau}\bar{\nu_l}$ is an irreducible background. However, the lepton momentum has a broad distribution for the background, while it depends only on the $\alpha$ mass for the signal and it appears as a bump over the irreducible background. 
In particular, we search for an excess over the normalized lepton energy spectrum $x_{\ell}$ of $\tau \to \ell\nu_{\tau}\bar{\nu_l}$, where $x_{\ell} = 2E^*_{\ell}/m_{\tau}$, performing template fits. 
The energy $E^*_{\ell}$ is defined in the approximate rest frame of $\tau_{\rm{sig}}$, i.e. where $E_{\tau} \approx \sqrt{s}/2$ and $\sqrt{s}$ is the energy in the center-of-mass frame, and the $\tau_{\rm{sig}}$ direction is opposite to the $\tau_{\rm{tag}}$ direction.
We do not find any significant excess in 62.8~fb$^{-1}$ of data, and we set world-leading 95\% C.L. upper limits to $\mathcal{B}(\tau \to \ell\alpha)/\mathcal{B}(\tau \to \ell\bar{\nu}_{\ell}\nu_{\tau})$, as a function of the $M_{\alpha}$ mass~\cite{taula}. Fig. \ref{fig:tau} shows the results for $l=e$.

\section{Summary}
We presented the latest Belle~II world-leading results from dark sector searches. The results use subsets of the 424~fb$^{-1}$ collected to date. New results with improved analyses and larger data samples  are expected to push further the Belle~II sensitivity to the dark sector.

\nocite{*}


\end{document}